\input wf20.sty
\rightline{\timestamp}
\brochureb{\smallsc the ground state
 of bottomium to two loops and higher}{\smallsc f. j.  yndur\'ain}{1}
\rightline{FTUAM 00-14}
\rightline{hep-ph/xxxxxxx}
\bigskip
\hrule height .3mm
\vskip.6cm
\centerline{{\twelverm The Ground State of Bottomium to Two Loops and Higher\footnote*{\petit 
This 
paper incorporates the results of the note ``Improved Determination of the $b$ Quark Mass 
From Spectroscopy", FTUAM--00--07, 2000 (hep--ph/0002237).\hb
\indent To be published in the Proceedings of the Seminar on Confinement, E. Schr\"odinger
 In ternational Institute, Vienna, May--June, 2000.}}}
\medskip
\centerrule{.7cm}
\vskip1cm

\setbox9=\vbox{\hsize65mm {\noindent\fib F. J. 
Yndur\'ain} 
\vskip .1cm
\noindent{\addressfont Departamento de F\'{\i}sica Te\'orica, C-XI,\hb
 Universidad Aut\'onoma de Madrid,\hb
 Canto Blanco,\hb
E-28049, Madrid, Spain.}\hb}
\smallskip
\centerline{\box9}
\bigskip
\setbox0=\vbox{\abstracttype{Abstract} We consider the properties
 of the ground state of bottomium. The 
$\upsilonv$ mass is evaluated to two loops, and including leading 
higher order [$O(\alpha_s^5\log\alpha_s)$] and $m_c^2/m_b^2$ corrections. 
This allows us to present  updated values for the pole mass and $\overline{MS}$ 
mass of the $b$ quark: $m_b=5022\pm58$ MeV, for the pole mass, 
and $\bar{m}_b(\bar{m}_b)=4286\pm36$ MeV for the $\overline{MS}$ one.
The value for the \msbar\ mass is accurate including 
 and $O(\alpha_s^3)$ corrections 
and leading orders in the ratio $m_c^2/m_b^2$.
We then consider the  
wave function for the ground state of  $\bar{b}b$, which is 
calculated to two loops in the nonrelativistic approximation. 
Taking into account the evaluation of the 
matching coefficients by Beneke and Signer one 
can calculate, in principle, 
 the width for the decay $\upsilonv\to e^+e^-$ to order $\alpha_s^5$. 
Unfortunately, given the size of the corrections it is impossible to 
produce reliable numbers. The situation is slightly 
better for the ground state of 
toponium, where a decay width into $e^+e^-$ of 11 -- 14 \kev\ is predicted.}
\centerline{\box0}

\brochureendcover{Typeset with \physmatex}

\brochuresection{1. Introduction}
In this paper we profit from the existence of recently obtained results
 on heavy quarkonium to give an accurate picture of the ground state of bottomium. 
In particular, we have the two loop relation between pole and \msbar\ mass\ref{1}; 
the $O(\alpha_s^5\log\alpha_s)$ and $O(m_c^2/m_b^2)$ corrections to 
quarkonium potential\ref{2}; 
and the two loop expression for the decay $\upsilonv\to e^+e^-$ 
in terms of the nonrelativistic ground state wave function\ref{3}, 
which last we evaluate here also to two loops. 

All this allows us to give a determination of the pole mass of the $b$ quark,
$$m_b=5022\pm58\;\mev$$
correct up to and including $\alpha_s^4$ and  $O(\alpha_s^5\log\alpha_s)$ corrections 
as well as leading nonperturbative and $O(m_c^2/m_b^2)$ ones; to find a 
value for the \msbar\ mass,
$$\bar{m}_b(\bar{m}_b)=4286\pm36\;\mev$$ 
correct to order $\alpha_s^3$ and leading nonperturbative and $O(m_c^2/m_b^2)$; 
and to present an evaluation of the width
$$\gammav(\upsilonv\to e^+e^-)$$
accurate to order $\alpha_s^5$ and including leading nonperturbative 
contributions. The determination of the 
pole mass turns out to be very stable, in particular when comparing it with older 
determinations\ref{4,5}. 
The \msbar\ mass is less stable, but on the other hand is in good 
agreement with recent NNLO (next-to-next to leading order) evaluations 
based on QCD sum rules\ref{6}. Finally, the 
values found for the width, while in rough 
agreement with experiment, show an excessive dependence
 both on the renormalization point chosen to 
calculate and on the degree of accuracy of the calculation: both due to the very 
large size of one and two loop corrections. 

The plan of this paper is as follows. In \sect~2 we present
 the hamiltonian used to obtain the results. 
This is used in \sect~3 to evaluate pole mass and \msbar\ 
mass for the $b$ quark, and to comment on the 
convergence properties of the series. 
The decay rates to $e^+e^-$ of the $\upsilonv$ are discussed in 
\sect~4. The article is finished 
with an Appendix where some of the 
evaluations, which may be of interest for other problems, are presented 
in detail.    

\brochuresection{2. The nonrelativistic  hamiltonian}
In the static limit the  $\bar{b}b$ interaction 
may be described by a potential. The ensuing 
hamiltonian may be written as
$$H=\widetilde{H}^{(0)}+H_1\equn{(2.1)}$$
where 
$$\widetilde{H}^{(0)}=2m+\dfrac{-1}{m}\lap-\dfrac{C_F\widetilde{\alpha}_s(\mu)}{r};\equn{(2.2a)}$$
we have put together all coulomb-like pieces of the interaction so that
$$\widetilde{\alpha}_s(\mu^2)=\alpha_s(\mu^2)
\left\{1+c^{(1)}\dfrac{\alpha_s(\mu^2)}{\pi}
+c^{(2)}\dfrac{\alpha_s^2}{\pi^2}\right\}\equn{(2.2b)}$$
$$\eqalign{c^{(1)}=&a_1+\dfrac{\gammae\beta_0}{2},\cr
c^{(2)}=&\gammae\left(a_1\beta_0+\dfrac{\beta_1}{8}\right)+
\left(\dfrac{\pi^2}{12}+\gammae^2\right)\dfrac{\beta_0^2}{4}+
a_2\cr}$$
and\ref{1,2}
$$\eqalign{a_1=&\dfrac{31C_A-20T_Fn_f}{36},\cr
a_2=&\tfrac{1}{16}
\Big\{\left[\tfrac{4343}{162}+4\pi^2-\tfrac{1}{4}\pi^4+\tfrac{22}{3}\zeta_3 \right]C_A^2\cr
&-\left[\tfrac{1798}{81}+\tfrac{56}{3}\zeta_3 \right]C_AT_Fn_f-
\left[\tfrac{55}{3}-16\zeta_3 \right]C_FT_Fn_f+\tfrac{400}{81}T_F^2n_f^2\Big\}\cr
&\simeq 13.2. \cr}\equn{(2.2c)} $$
The remaining piece of the interaction $H_1$ is, in the non-relativistic (static) limit, 

$$H^{\rm NR}_1=V^{(L,1)}+V^{(L,2)}+V^{(LL)},\equn{(2.3a)}$$
$$\eqalign{V^{(L,1)}=&
\dfrac{-C_F\alpha_s(\mu^2)^2}{\pi}\,\dfrac{\beta_0}{2}\dfrac{\log r\mu}{r},\cr
V^{(L,2)}=&
\dfrac{-C_F\alpha_s^3}{\pi^2}\,
\left(a_1\beta_0+\dfrac{\beta_1}{8}+\dfrac{\gammae\beta_0^2}{2}\right)\dfrac{\log r\mu}{r},\cr
V^{(LL)}=&
\dfrac{-C_F\beta_0^2\alpha_s^3}{4\pi^2}\,\dfrac{\log^2 r\mu}{r}.\cr}
\equn{(2.3b)}$$
 Note that the mass 
that appears in above formulas is the {\sl pole} mass, but all
 other renormalization is carried in the \msbar\ scheme. 
For more details on this see refs.~4,~5.

\equs{(2.2)} give what we will need to calculate 
the wave function, when taking into account also the results of Beneke et al.\ref{3}; 
but for the spectrum we can do better. First, we have to add one loop velocity corrections. 
For the spin-independent part of the spectrum this is
$$H_{\rm vel.\,dep.}=
\dfrac{C_Fb_1\alpha_s^2}{2mr^2},\equn{(2.4a)}$$
$b_1=\tfrac{1}{2}(C_F-2C_A)$
and a spin-dependent piece
$$V_{\rm hf}=
\dfrac{4\pi C_F\alpha_s}{3m^2}s(s+1)\delta({\bf r})\equn{(2.4b)}$$
 has to be added for vector states like the $\upsilonv$. 
Then we 
have the corrections yielding contributions of order 
$\alpha_s^5\log\alpha_s$, which are the leading $O(\alpha_s^5)$ corrections 
in the limit of small $\alpha_s$. 
They have been evaluated in ref.~2 
and they lead to a correction on the mass of the $\upsilon$ of 
$$\delta_{[\alpha_s^5\log\alpha_s]} E_{10}=
-m[C_F+\tfrac{3}{2}C_A]C^4_F\alpha_s^5(\log\alpha_s)/\pi,\equn{(2.5)}$$
for $\mu=2/a=mC_F\alpha_s$.  
 
The influence of the nonzero mass of the $c$ quark, the only worth considering, 
will be evaluated now. 
To leading order it only contributes to the $\bar{b}b$ potential 
through a $c$-quark loop in the gluon exchange 
diagram. The momentum space 
potential generated by a nonzero mass quark through this 
mechanism is then, in the nonrelativistic limit,
$$\widetilde{V}_{c\,{\rm mass}}=-\dfrac{8C_FT_F\alpha_s^2}{{\bf k}^2}
\int_0^1\dd x\,x(1-x)\log\dfrac{m_c^2+x(1-x){\bf k}^2}{\mu^2}.\equn{(2.6)} $$
We expand in powers of $m_c^2/{\bf k}^2$. The zeroth term 
is already included in (5). The first order correction is\ref{2}
$$\delta_{c\,{\rm mass}}\widetilde{V}=-\dfrac{8C_FT_F\alpha^2_sm_c^2}{{\bf k}^4}.
 \equn{(2.7)} $$
In x-space,
$$\delta_{c\,{\rm mass}} V=\dfrac{C_FT_F\alpha_s^2m_c^2}{\pi}\,r.\equn{(2.8)} $$ 

If we are interested in a calculation correct to order 
$\alpha_s^4$ and $\alpha_s^5\log\alpha_s$ for the bound state 
energies, or order $\alpha_s^5$ for the decay rate, 
all these terms have to be treated as first order perturbations except 
$V^{(L,1)}$ that should be computed to second order.\fnote{In 
fact, this is the most difficult part of the calculation.} Of course the hamiltonian 
$\widetilde{H}^{(0)}$  can, and will, be solved exactly.
 
\brochuresection{3. The $b$ quark mass}
\brochuresubsection{3.1. Pole mass}
Here $\alpha_s$ is to be calculated to three loops:
$$\alpha_s(\mu)=\dfrac{4\pi}{\beta_0L}\left\{1-\dfrac{\beta_1\log L}{\beta_0^2L}+
\dfrac{\beta_1^2\log^2L-\beta_1^2\log L+\beta_2\beta_0-\beta_1^2}{\beta_0^4L^2}\right\}
\equn{(3.1)}$$
with
$$L=\log\dfrac{\mu^2}{\Lambdav^2};\quad
\beta_0=11-\tfrac{2}{3}n_f,\quad
\beta_1=102-\tfrac{38}{3}n_f,\quad
\beta_2=\tfrac{2857}{2}-\tfrac{5033}{18}n_f+\tfrac{325}{54}n_f^2.$$
We take as input parameters
$$\Lambdav(n_f=4,\,\hbox{three loops})=0.283\pm0.035\;\gev\;
\left[\;\alpha_s(M_Z^2)\simeq0.117\pm0.024\;\right]
$$ 
(ref.~7) and for the gluon condensate, very poorly known, the value 
$\langle\alpha_sG^2\rangle=0.06\pm0.02\;\gev^4.$
From the mass of the $\upsilonv$ particle we 
have a very precise determination for the pole mass of the $b$ quark. 
This determination is correct to order $\alpha_s^4$ 
and including leading $O(m_c^2/m_b^2)$ and 
leading nonperturbative corrections as well as the 
$\alpha_s^5$ corrections proportional to  $\log\alpha_s$; 
the details of it will be given below. With 
 the renormalization point 
$\mu=m_bC_F\alpha_s$ we have,
$$\eqalign{m_b=&
5022\pm43\,(\Lambdav)\;\mp5\,(\langle\alpha_sG^2\rangle)^{-31}_{+37} \;
(\hbox{vary}\; \mu^2\;{\rm by}\,25\%)
\;\pm 38\;({\rm other\; th.\;uncert.})\cr
=&5022\pm58\;\mev.}
\equn{(3.2)}$$
Here we append $(\lambdav)$ to the error 
induced by that of $\lambdav$, and likewise 
$(\langle\alpha_sG^2\rangle)$ tags the error due to that of the condensate. 
The error labeled (other th. uncert.) includes also the error evaluated in ref.~8; 
the rest is as in ref.~5.

We collect in the table the determinations of the $b$ quark mass 
based on spectroscopy, to increasing accuracy. 
The {\sl stability} of the numerical values of the pole mass is 
remarkable: 
the pole masses  all lie within each other error bars.
 The \msbar\ ones 
(that will be discussed in greater detail in next subsection) show more spread. It may be noted that the three loop value for 
$\bar{m}_b(\bar{m}_b^2)$, $4286\pm36$, agrees 
perfectly with the Beneke--Signer three loop 
result\ref{6} based on sum rules, $\bar{m}_b(\bar{m}_b^2)=4260\pm100\,\mev$. 
This agreement, however, is less satisfactory than it would appear at first sight 
because of the large size of the $O(\alpha_s^3)$ corrections. 
We will discuss this more in next subsection. 
\bigskip
\setbox0=\vbox{
\setbox1=\vbox{\offinterlineskip\hrule
\halign{
&\vrule#&\strut\hfil#\hfil&\vrule#&\strut\quad#\quad&\vrule#&\strut\quad#\quad&\vrule#&\strut\quad#\quad&\vrule#&\strut\quad#\cr
 height2mm&\omit&&\omit&&\omit&&\omit&&\omit&\cr 
&\kern0.2em Reference\kern0.2em&&$m_b({\rm pole})$&&$\bar{m}_b(\bar{m}_b^2)$&&$m_c({\rm pole})$&& $\bar{m}_c(\bar{m}_c^2)$& \cr
 height1mm&\omit&&\omit&&\omit&&\omit&&\omit &\cr
\noalign{\hrule} 
height1mm&\omit&&\omit&&\omit&&\omit&&\omit&\cr
&TY&& $4971\pm72$&&$4401^{+21}_{-35}$\vphantom{$4^{4^4}_{4_4}$}&&
$1585\pm 20\,(^*)$&&$1321\pm 30\,(^*)$\phantom{\big{]}}&\cr
&PY&& $5065\pm60$&&$4455^{+45}_{-29}$\vphantom{$4^{4^4}_{4_4}$}&&
$1866^{+215}_{-133}$&&$1542^{+163}_{-104}$&\cr
&Here&& $5022\pm58$&&$4286\pm36$&&
$-$&&$-$\vphantom{\big\}}&\cr
 height1mm&\omit&&\omit&&\omit&&\omit&&\omit&\cr
\noalign{\hrule}}
\vskip.05cm}
\centerline{\box1}
{\petit
\centerline{{\sc Table 1.} $b$ and $c$ quark masses.\quad $(^*)$
 Systematic errors not included.}}
\vskip-0.2cm
\centerrule{0.3cm}
\smallskip
\setbox2=\vbox{\hsize=0.95\hsize 
\petit{
\noindent
TY: Titard and Yndur\'ain\ref{4}. $O(\alpha_s^3)$ plus $O(\alpha_s^3)v$, $O(v^2)$ 
for $m$;
$O(\alpha_s^2)$ for $\bar{m}$. 
 Rescaled for 
$\lambdav(n_f=4)=283\,\mev$.\hb
PY: Pineda and Yndur\'ain\ref{5}. Full $O(\alpha_s^4)$ for $m$; 
$O(\alpha_s^2)$ for $\bar{m}$. Rescaled for 
$\lambdav(n_f=4)=283\,\mev$.\hb
Here: This calculation. $O(\alpha_s^4)$, $O(\alpha_s m_c^2/m_b^2)$ 
and $O(\alpha_s^5\log\alpha_s)$ 
 for $m$; $O(\alpha_s^3)$ and  $O(\alpha_s^2 m_c^2/m_b^2)$  for  $\bar{m}$. 
Values not given for the $c$ 
quark, as the higher order terms are as large as the leading ones.}}
\centerline{\box2}}
\centerline{\box0}
\centerrule{0.3cm}
\smallskip
We finally remark that the values of $m_b$ quoted e.g. in 
the Table~1 were {\sl not} obtained solving \equn{(8)}, but 
solving exactly the coulombic part of the interaction,
 and perturbing the result (see refs.~4,~5 for details). 
We also note that, in the determinations of $m_b$, the new pieces,
 $O(\alpha_s m_c^2/m_b^2)$ 
and $O(\alpha_s^5\log\alpha_s)$, have been evaluated to first order; 
in particular, we have 
included the corresponding shifts in the {\sl central} values, not in the errors. 
If we included these, the errors would decrease by some 7\%. 

\brochuresubsection{3.2. $m_b\,-\,\bar{m}_b(\bar{m}_b) $ connection} 
The connection of the pole mass with the \msbar\ 
mass has been known for some time to one and two loops: very 
recently, a three loop evaluation has been completed. 
Coupling this with the pole mass evaluations, we now have 
an order $\alpha_s^3$ result for the \msbar\ mass. 
We review here briefly this.

Write, for a heavy quark,  
$$\bar{m}(\bar{m})\equiv m/\{1+\delta_1+\delta_2+\delta_3+\cdots\};\equn{(3.3a)}$$
$m$ here denotes the {\sl pole} mass, and $\bar{m}$ is  the \msbar\ one. 
One has
$$\delta_1=C_F\dfrac{\alpha_s(\bar{m})}{\pi},\quad
\delta_2=c_2\left(\dfrac{\alpha_s(\bar{m})}{\pi}\right)^2,\quad
\delta_3=c_3\left(\dfrac{\alpha_s(\bar{m})}{\pi}\right)^3.\equn{(3.3b)}$$

The coefficient $c_2$ has been evaluated by Gray et al.\ref{1}, and reads
$$c_2=-K+2C_F,\equn{(3.3c)}$$
$$\eqalign{K=&\,K_0+\sum_{i=1}^{n_f}\Deltav\left(\dfrac{m_i}{m}\right),\quad
K_0=\tfrac{1}{9}\pi^2\log2+\tfrac{7}{18}\pi^2
-\tfrac{1}{6}\zeta(3)+\tfrac{3673}{288}-\left(\tfrac{1}{18}\pi^2+
\tfrac{71}{144}\right)(n_f+1)\cr
\simeq&\,16.11-1.04\,n_f;\quad 
\Deltav(\rho)=\tfrac{4}{3}\left[\tfrac{1}{8}\pi^2\rho-\tfrac{3}{4}\rho^2+\cdots\right].
 \cr}\equn{(3.3d)}$$
$m_i$ are the (pole) masses of the quarks strictly lighter 
than $m$, and $n_f$ is the number of these. For the $b$ quark case, $n_f=4$ 
and only the $c$ quark mass has to be considered; we will take $m_c=1.8\,\gev$ 
(see Table 1 above)  
for the calculations.

The coefficient $c_3$ was recently calculated by Melnikov and van Ritbergen\ref{1}. 
Neglecting now the $m_i$,
$$c_3\simeq                                  
190.389 - 26.6551 n_f + 0.652694 n_f^2. \equn{(3.3e)}$$

For the $b$, $c$ quarks, with $\alpha_s$ as given before,
$$\eqalign{\delta_1(b)=&0.090,\cr
\delta_2(b)=&0.045,\cr
\delta_3(b)=&0.029;\cr}\quad
\eqalign{\delta_1(c)=&0.137,\cr
\delta_2(c)=&0.108,\cr
\delta_3(c)=&0.125.\cr}\equn{(3.4)}$$

From these values we conclude that, for the $c$ quark, 
the series has started to diverge at second order, and it certainly 
diverges at order $\alpha_s^3$. For the $b$ quark
 the series is at the edge of convergence 
for the $\alpha_s^3$ contribution.

Using the three loop relation (3.3) of the pole mass to the \msbar\ mass 
we then find, from the results in \subsect~3.1 for the pole mass, the value
$$\bar{m}_b(\bar{m}_b)=4284\pm7\;(\lambdav)\mp5\;(\langle\alpha_s G^2\rangle)
\pm35\;(\hbox{other th. uncert.})=4284\pm36\;\mev.\equn{(3.5)}$$
This is the value incorporated in Table~1. 
The slight dependence 
of $\bar{m}$ on $\lambdav$ when evaluated in this way was already noted in 
ref.~4.

There is another way of obtaining $\bar{m}$, which is to express directly 
the mass of the $\upsilonv$ in terms of it, using
 \equn{(3.3)} and the order $\alpha_s^3$ formula for the $\upsilonv$ 
mass in terms of the pole mass \equn{(3.3a)}. One finds, for $n_f=4$, 
and neglecting $m_c^2/m_b^2$,
$$\eqalign{M(\upsilonv)=&2\bar{m}(\bar{m})
\Big\{1+C_F\dfrac{\alpha_s(\bar{m})}{\pi}+7.559\left(\dfrac{\alpha_s(\bar{m})}{\pi}\right)^2\cr
+&\Big[66.769+18.277\left(\log C_F+
\log\alpha_s(\bar{m})\right)\Big]
\left(\dfrac{\alpha_s(\bar{m})}{\pi}\right)^3\Big\}.\cr}
\equn{(3.6a)}$$
(One could add the leading nonperturbative contributions to (3.6a) 
\`a la Leutwyler--Voloshin in the standard way).
This method has been at times advertised as improving the 
convergence, allegedly because the \msbar\ mass does not 
suffer from nearby renormalon singularities. 
But a close look to (3.6a) does not seem to bear this out. 
To an acceptable $O(\alpha_s^4)$ error we can replace 
$\log(\alpha_s(\bar{m}))$ by $\log(\alpha_s(M(\upsilonv/2))$ above.  
With $\lambdav$ as before  
(3.6a) then becomes
$$M(\upsilonv)=2\bar{m}_b(\bar{m}_b)
\left\{1+C_F\dfrac{\alpha_s(\bar{m})}{\pi}
+7.559\left(\dfrac{\alpha_s(\bar{m})}{\pi}\right)^2
+43.502\left(\dfrac{\alpha_s}{\pi}\right)^3\right\}.
\equn{(3.6b)}$$
This does not look particularly convergent, and is certainly not 
an  
improvement over the expression using the pole mass where one has, 
for the choice $\mu=C_Fm_b\alpha_s$ and 
still neglecting the masses of quarks lighter 
than the $b$, 
$$M(\upsilonv)=2m_b\left\{1-2.193\left(\dfrac{\alpha_s(\mu)}{\pi}\right)^2-
24.725\left(\dfrac{\alpha_s(\mu)}{\pi}\right)^3-
458.28\left(\dfrac{\alpha_s(\mu)}{\pi}\right)^4+
897.93\,[\log\alpha_s] \left(\dfrac{\alpha_s}{\pi}\right)^5\right\}.\equn{(3.7)}$$
To order three, (3.7) is actually better\fnote{The convergence of \equn{(3.7)} 
is still improved if one solves exactly the purely coulombic part of the 
static potential, as was done in \subsect~3.1. 
For example, the $O(\alpha_s^4)$ term would become
$-232.12(\alpha_s/\pi)^4$.} 
than (3.6b). What is more, logarithmic 
terms appear in (3.6) at order $\alpha_s^3$, 
while for the pole mass expression they first show up at order $\alpha_s^5$.    
Finally, the direct formula for $M(\upsilonv)$ in terms of the \msbar\ 
mass presents the extra difficulty that the {\sl nonperturbative} 
contribution becomes  larger than than what one has
 for the expression in terms of the pole mass 
($\sim 80$ against $\sim9$ \mev), 
because of the definition of the renormalization point. 
With the purely perturbative expression (3.6) plus leading 
nonperturbative (gluon condensate) correction one finds the 
value $\bar{m}_b(\bar{m}_b)=4167\mev$, 
rather low.

\brochuresection{4. The Decay Rates $\upsilonv\to e^+e^-$ and $T\to e^+e^-$ }
For bound state calculations it is convenient to 
solve $\widetilde{H}^{(0)}$ exactly, and treat 
$H_1$ as a perturbation; but, to evaluate 
the wave function it is preferable to work in the following 
somewhat different manner. 
Define the quantities
$$a\equiv\dfrac{2}{mC_F\alpha_s(\mu)},\quad \rho\equiv \dfrac{2r}{a}.$$
One may rewrite $H_1$ as
$$H_1=H_{1C}+H^{(1)}_{1L}+H^{(2)}_{1L}+H_{1LL};
$$
$$\eqalign{H_{1C}=&\left(\log\dfrac{\mu a}{2}\right)
\left\{\tfrac{1}{2}\beta_0\dfrac{\alpha_s(\mu)}{\pi}+
\left[c_L+\dfrac{\beta_0^2\log\mu a/2}{4}\right]\dfrac{\alpha_s^2(\mu)}{\pi^2}\right\}
\dfrac{-C_F\alpha_s(\mu)}{r}\cr
H^{(1)}_{1L}=&-\dfrac{2}{a}\,\dfrac{C_F\beta_0\alpha_s^2(\mu)}{2\pi}
\,\dfrac{\log\rho}{\rho};\cr
H^{(2)}_{1L}=&-\dfrac{2}{a}\,\left[ c_L+\dfrac{\beta_0^2}{2}\log\dfrac{\mu a}{2}\right]
\dfrac{C_F\alpha_s^3(\mu)}{\pi^2}\,\dfrac{\log\rho}{\rho};\cr
H_{1LL}=&-\dfrac{2}{a}\,\,\dfrac{C_F\beta_0^2\alpha_s^3}{4\pi^2}\,\dfrac{\log\rho}{\rho}.
\cr}
\equn{(4.1)}$$
Here $c_L=a_1\beta_0+\tfrac{1}{8}\beta_1+\gammae\beta_0^2/2$.

One can put $\widetilde{H}^{(0)}$ and $H_{1C}$ together, 
$\bar{H}^{(0)}=\widetilde{H}^{(0)}+H_{1C}$, and solve the 
corresponding coulombic Schr\"odinger equation exactly by just replacing, in the 
ordinary solution of the coulombic problem,
$$\eqalign{H^{(0)}\psiv_{nl}^{(0)}({\bf r})=&E^{(0)}_n\psiv_{nl}^{(0)}({\bf r}),\cr
H^{(0)}=&-\dfrac{1}{m}\lap-\dfrac{C_F\alpha_s}{r},\cr
\psiv_{nl}^{(0)}({\bf r})=&\dfrac{1}{\sqrt{4\pi}}R_{nl}^{(0)}(r),\quad
 R_{10}^{(0)}(r)=\alpha_s^{3/2}\dfrac{(mC_F)^{3/2}}{\sqrt{2}}\ee^{-rC_Fm\alpha_s/2},\cr}
\equn{(4.2)}$$
the coupling $\alpha_s$ according to
$$\alpha_s\to\alpha_s\left\{1+c^{(1)}\dfrac{\alpha_s(\mu^2)}{\pi}
+c^{(2)}\dfrac{\alpha_s^2}{\pi^2}+
\left(\log\dfrac{\mu a}{2}\right)
\left[\tfrac{1}{2}\beta_0\dfrac{\alpha_s(\mu)}{\pi}+
\left(c_L+\dfrac{\beta_0^2\log\mu a/2}{4}\right)\dfrac{\alpha_s^2(\mu)}{\pi^2}\right]\right\}.$$
The remaining $H^{(2)}_{1L}+H_{1LL}$ can then be evaluated as first order perturbations; only  
 $H^{(1)}_{1L}$ has to be calculated to first and second order and, for this second order, 
the fact that one perturbs on $\bar{H}^{(0)}$ has also to be taken into account. 

\brochuresubsection{4.1 Calculation of perturbations}
The NNLO (two loop) hard part of the radiative correction to the 
leptonic decay of quarkonium, say
$$\gammav(\upsilonv\to e^+e^-),$$
has been evaluated some time ago\ref{3}. One can then write
$$\gammav(\upsilonv\to e^+e^-)=\left[\dfrac{Q_b\alpha_{\rm QED}}{M(\Upsilonv)}\right]
\left|\left\{1+\delta^{(1)}_{\rm hard}\dfrac{\alpha_s(\mu)}{\pi}+
\delta^{(2)}_{\rm hard}\dfrac{\alpha^2_s(\mu)}{\pi^2}
\right\}R_{10}(0)\right|^2,\equn{(4.3a)}$$
where\ref{1}

$$\eqalign{\delta^{(1)}_{\rm hard}=&-2C_F,\cr
\delta^{(2)}_{\rm hard}=&C_F^2\left\{\pi^2\left[\tfrac{1}{6}\log\dfrac{m^2}{\mu^2}
-\tfrac{79}{36}+\log2\right]
+\tfrac{23}{8}-\tfrac{1}{2}\zeta_3\right\}\cr
+&C_FC_A\left\{\pi^2\left[\tfrac{1}{4}\log\dfrac{m^2}{\mu^2}
+\tfrac{89}{144}-\tfrac{5}{6}\log2\right]-\tfrac{151}{72}-
\tfrac{13}{4}\zeta_3\right\}\cr
+&\tfrac{11}{18}C_FT_Fn_f+C_FT_F\left(-\tfrac{2}{9}\pi^2+\tfrac{22}{9}\right)\cr
+&\dfrac{C_F\beta_0\log m^2/\mu^2}{2},
\cr}\equn{(4.3b)}$$
and $R_{10}(0)$ is the static (radial) wave function, i.e., evaluated neglecting 
terms of relative order $1/m$ in the interaction. 
We will write
$$R_{10}(0)=\left\{1+\delta^{(1)}_{\rm wf}\dfrac{\alpha_s(\mu)}{\pi}+
\delta^{(2)}_{\rm wf}\dfrac{\alpha^2_s(\mu)}{\pi^2}
+
\delta_{\rm NP}\right\}R^{(0)}_{10}(0);
\equn{(4.3c)}$$
$R^{(0)}_{10}(0)$ is as in (4.2) and the $\delta_{\rm wf}$ are to be evaluated with 
$H$ in (2.1) (or (4.1)). Finally,  the (leading) 
nonperturbative piece $\delta_{\rm NP}$ may be 
obtained in terms of the gluon condensate $\langle \alpha_s G^2\rangle$
 as in refs.~4,~9 so that
$$\delta_{\rm NP}=
\dfrac{2\,968\pi\langle \alpha_s G^2\rangle}{425m^4(C_F\widetilde{\alpha}_s)^6}.
\equn{(4.4)}$$
Define then,  with  self-explanatory notation,
$$\delta_{\rm wf}^{(1)}\dfrac{\alpha_s}{\pi}+
\delta_{\rm wf}^{(2)}\left(\dfrac{\alpha_s}{\pi}\right)^2=
\deltav_C+\deltav^{(1)}_{L,1}+\deltav^{(1)}_{L,2}+\deltav^{(1)}_{LL}+
\deltav^{(2)}_{L,1}.\equn{(4.5)}$$ 
$\deltav_{C}$ is calculated trivially, as it is equivalent to a 
modification of the coulombic potential in $H^{(0)}$. 
The remaining $\delta$s are easily evaluated with the formulas of the Appendix. 
We find, 
$$\eqalign{\deltav_{C}=&\tfrac{3}{2}\left[c_1+\dfrac{\beta_0\log\mu a/2}{2}\right]
\dfrac{\alpha_s(\mu)}{\pi}\cr
+&\Big\{\tfrac{3}{8}\left[c_1+\dfrac{\beta_0\log\mu a/2}{2}\right]^2
+\tfrac{3}{2}\left[c_2
+2c_L\log\dfrac{a\mu}{2}+\dfrac{\beta_0^2\log^2a\mu/2}{4}\right]
\dfrac{\alpha_s^2}{\pi^2}\Bigg\};\cr}
\equn{(4.6a)}$$
$$\eqalign{\deltav^{(1)}_{L,1}=&-\dfrac{3\beta_0\alpha_s(\mu)}{4\pi}
\left(\gammae+\dfrac{\pi^2-6}{9}\right)
+3\beta_0\left(\gammae+\dfrac{\pi^2-6}{9}\right)\left(c^{(1)}+\dfrac{\beta_0\log a\mu/2}{2}\right)
\dfrac{\alpha^2_s(\mu)}{4\pi^2};\cr
\deltav^{(1)}_{L,2}=&-3\left(\gammae+\dfrac{\pi^2-6}{9}\right)
\left[2c_L+\beta_0^2\log\dfrac{a\mu}{2}\right]\dfrac{\alpha^2_s(\mu)}{4\pi^2};\cr
\deltav^{(1)}_{LL}=&
\beta_0^2\left\{2\zeta_3-(1-\gammae)\left(1+\dfrac{\pi^2}{3}\right)
+\tfrac{3}{2}\left[\gammae^2-2\gammae+\dfrac{\pi^2}{6}\right]\right\}
\dfrac{\alpha^2_s(\mu)}{4\pi},
\cr}\equn{(4.6b)}$$
where $\deltav_{L,1}^{1)}$ above actually 
contains the NNLO mixed coulombic-logarithmic correction.
Finally,
$$\deltav^{(2)}_{L,1}=c_{L,1}^{(2)}\dfrac{\beta_0^2\alpha_s^2}{4\pi^2},\qquad
c_{L,1}^{(2)}\simeq1.75.
\equn{(4.6c)}$$
For the exact value of $c_{L,1}^{(2)}$, see the Appendix, \equs~(A9). 
This finishes the calculation of the decay rate.

\booksubsection{4.2 NLO calculation}
Using (4.1-6) to one loop only we find the 
NLO result,
$$\eqalign{\gammav(\upsilonv\to e^+e^-)=&
\left\{1+\left[\dfrac{3\beta_0}{4}\left(\log\dfrac{a\mu}{2}-
\gammae+\dfrac{6-\pi^2}{9}\right)+\tfrac{3}{2}\left(\dfrac{\gammae\beta_0}{2}+
\dfrac{93-10n_f}{36}\right)
 -2C_F\right]\dfrac{\alpha_s}{\pi}\right\}\cr
\times&\gammav^{(0)}(\upsilonv\to e^+e^-)\cr}
\equn{(4.7)}$$
where the LO expression $\gammav^{(0)}$ is 
$$\gammav^{(0)}=2\left[\dfrac{Q_b\alpha_{\rm QED}}{M(\Upsilonv)}\right]
[mC_F\alpha_s(\mu)]^3.$$

We note that (4.7) is slightly different from the result obtained using a variational method 
(instead of Rayleigh--Schr\"odinger perturbation theory) to evaluate
 $\delta R_{10}$, 
which gives
$$\gammav(\upsilonv\to e^+e^-)=\left\{1+\left[\dfrac{3\beta_0}{4}\left(\log\dfrac{a\mu}{2}-
\gammae\right)
+\tfrac{3}{2}\left(\dfrac{\gammae\beta_0}{2}+
\dfrac{93-10n_f}{36}\right) -2C_F\right]\dfrac{\alpha_s}{\pi}\right\}
\gammav^{(0)}(\upsilonv\to e^+e^-).
$$

\booksubsection{4.3 NNLO calculation, and numerical results}
The NNLO result is obtained with the full (4.3) -- (4.6) expressions, with which 
numerical results are readily obtained. We select the values of the 
basic parameters
$\Lambdav(n_f=4,\,\hbox{three loops})$,  
$\langle\alpha_sG^2\rangle$ as before.
For the renormalization point there are two ``natural" choices:  
$$\mu_1=m_b\quad{\rm and}\quad\mu_2=2/a=2.866\,\gev.$$
The first value gives a reasonable value for the width, 
These are the input values taken for the calculations reported in the Table 2 below.
\bigskip
\setbox0=\vbox{
\setbox1=\vbox{\offinterlineskip\hrule
\halign{
&\vrule#&\strut\hfil#\hfil&\vrule#&\strut\quad#\quad&\vrule#&\strut\quad#\quad&\vrule#&\strut\quad#\quad&\vrule#&\strut\quad#&#\cr
 height2mm&\omit&&\omit&&\omit&&\omit&&\omit&\cr 
&\kern1.5mm$\Gammav(\Upsilonv\to e^+e^-)$\kern1.5mm&
&\hfil LO\hfil&&\hfil NLO\hfil&&\hfil NNLO\hfil& \cr
 height1mm&\omit&&\omit&&\omit&&\omit&\cr
\noalign{\hrule} 
height1mm&\omit&&\omit&&\omit&&\omit&\cr
&$\mu=m$&& $0.41$&&$1.22$\vphantom{$4^{4^4}_{4_4}$}&&
$1.13$ \kev&\cr
&$\mu=2/a$&& $0.73$&&$0.55$\vphantom{$4^{4^4}_{4_4}$}&&
$0.80$ \kev&\cr
 height1mm&\omit&&\omit&&\omit&&\omit&\cr
\noalign{\hrule}}
\vskip.05cm}
\centerline{\box1}

\smallskip
\setbox2=\vbox{\hsize=0.95\hsize 
\centerline{\petit{
\noindent {\sc Table 2}: Determination of $\gammav(\upsilonv\to e^+e^-)$ to increasing 
accuracy. }}}
\centerline{\box2}}
\centerline{\box0}
\centerrule{3cm}
\smallskip
\noindent The experimental figure is
$$\Gammav(\Upsilonv\to e^+e^-)=1.32\pm0.04\,\kev.$$
Clearly,  the large  NLO and NNLO perturbative corrections, 
both of similar size, and 
of  the leading NP correction,   
 make the theoretical result unstable, as the Table shows.
 
For the (perhaps measurable) 
toponium (T) width we  get, for $m_t=175\,\gev$, the corresponding results, 
summarized in Table 3.

\bigskip
\setbox9=\vbox{
\setbox1=\vbox{\offinterlineskip\hrule
\halign{
&\vrule#&\strut\hfil#\hfil&\vrule#&\strut\quad#\quad&\vrule#&\strut\quad#\quad&\vrule#&\strut\quad#\quad&\vrule#&\strut\quad#&#\cr
 height2mm&\omit&&\omit&&\omit&&\omit&&\omit&\cr 
&\kern0.5mm$\Gammav(T\to e^+e^-)$\kern0.5mm&
&\hfil LO\hfil&&\hfil NLO\hfil&&\hfil NNLO\hfil& \cr
 height1mm&\omit&&\omit&&\omit&&\omit&\cr
\noalign{\hrule} 
height1mm&\omit&&\omit&&\omit&&\omit&\cr
&$\mu=m_t$&& $6.86$&&$10.53$\vphantom{$4^{4^4}_{4_4}$}&&
$13.0$&\cr
&$\mu=2/a$&& $10.24$&&$10.91$\vphantom{$4^{4^4}_{4_4}$}&&
$13.5$&\cr
 height1mm&\omit&&\omit&&\omit&&\omit&\cr
\noalign{\hrule}}
\vskip.05cm}
\centerline{\box1}
\smallskip
\setbox2=\vbox{\hsize=0.95\hsize 
\centerline{\petit
\noindent {\sc Table 3}: Determination of $\gammav(T\to e^+e^-)$ to increasing 
accuracy.}}
\centerline{\box2}
\centerrule{3cm}
\smallskip}
\centerline{\box9}

The situation has improved with respect to what we had 
for bottomium, but the dependence on the renormalization point and on the 
order of perturbation theory considered is still a bit large. 
We can however conclude on an {\sl estimate} of some 
$11.0\,-\,14$ \kev\ for the width.

\brochuresection{Appendix: Perturbation by terms $\rho^{-1}\log^n\rho$}
To obtain the result of perturbations by terms $\rho^{-1}\log^n\rho$ we 
first evaluate the perturbation by a term $\rho^\nu$, $\nu=$~integer, 
continue to noninteger $\nu$ and differentiate with respect to $\nu$ at 
$\nu=-1$. For this we use the method devised by Leutwyler\ref{5} 
and extended 
in the second paper of ref~5, Appendix, in general. 
The basic formula is the inversion formula
$$\dfrac{1}{E_1^{(0)}-H^{(0)}(\kappa)} \rho^\mu R_{10}(\rho)=
\dfrac{ma^2}{4}\dfrac{\gammav(\mu+2)\gammav(\mu+3)}{\gammav(\mu+3-\kappa)}
\left(\sum_{j=0}^{\mu+1}\dfrac{\gammav(j+1-\kappa)}{\gammav(j+1)\gammav(j+2)}\rho^j\right)
 R_{10}(\rho);$$
here
$$H(\kappa)=H_0+\dfrac{\kappa C_F\alpha_s}{r},\qquad  R_{10}(\rho)=\dfrac{2}{a^{3/2}}\ee^{-\rho/a}$$
with $H_0$ the free hamiltonian and $a=2/mC_F\alpha_s$. This 
inversion formula is
valid for any $\kappa$ and integer $\mu$.
Using this one may then check that the first order 
perturbation by $\rho^\nu$, $\delta_{(\nu)} R_{10}$, is
$$\eqalign{\delta_{(\nu)} R_{10}(\rho)=
P_{10}\dfrac{1}{E_1^{(0)}-H^{(0)}}P_{10}\rho^\nu R_{10}&=
-\dfrac{ma^2}{4}\gammav(\nu+3)\cr
\times\Bigg\{\tfrac{3}{2}-&\tfrac{1}{2}\rho-\left[\dfrac{\nu+1}{2}+\psi(\nu+2)-\psi(1)\right]+
\sum_{j=1}^{\nu+1}\dfrac{\rho^j}{j\gammav(j+2)}\Bigg\} R_{10}(\rho).
\cr}
\equn{(A1)}$$
Sums like that in (A1) may be continued to arbitrary 
$\nu$ by using the formula
$$\sum_{j=1}^{\nu+1}f(j)=\sum_{j=1}^\infty\left[f(j)-f(j+\nu+1)\right]. 
\equn{(A2)}$$
This replacement is valid provided the sum $\sum_{j=1}^{\infty}f(j)$ is 
convergent; otherwise, we have to separate from $f$ the leading, next to leading\tdots pieces, 
to be summed explicitely, leaving a residue for which the sum up to infinity converges. 
This has already been done in getting (A1). 
As a check of this, we mention that the result
 agrees, for $\nu=-1$, with the result of the 
(trivial) evaluation obtained directly by replacing the coulombic 
potential according to 
$$\dfrac{\kappa C_F\alpha_s}{r}\to\dfrac{\kappa C_F\alpha_s}{r}+\dfrac{1}{\rho}$$
and expanding: $\delta_{(-1)}R_{10}(\rho)=-(ma^2/4)(\tfrac{3}{2}-\tfrac{1}{2}\rho)R_{10}(\rho)$. 
The results of first order perturbation by $\rho^{-1}\log \rho$, 
 $\rho^{-1}\log^2 \rho$ are then found by differentiating (A1) with respect to $\nu$.

To second order we require
$$\bar{\delta}^2_{(\lambda\nu)} R_{10}=\delta^2_{1(\lambda\nu)} R_{10}-
\delta^2_{2(\lambda\nu)} R_{10},$$
$$\delta^2_{1(\lambda\nu)} R_{10}=
P_{10}\dfrac{1}{E_1^{(0)}-H^{(0)}}P_{10}\rho^\lambda
P_{10}\dfrac{1}{E_1^{(0)}-H^{(0)}}P_{10}\rho^\nu R_{10}$$
and
$$\delta^2_{2(\lambda\nu)} R_{10}=\langle R_{10}|\rho^\mu|R_{10}\rangle\,
P_{10}\dfrac{1}{E_1^{(0)}-H^{(0)}}
P_{10}\dfrac{1}{E_1^{(0)}-H^{(0)}}P_{10}\rho^\nu R_{10}.$$
Here $P_{10}$ is the projector $P_{10}=1-|R_{10}\rangle\langle R_{10}|$. 
At $\rho=0$ we then have,
$$\eqalign{\delta^2_{1(\lambda\nu)} R_{10}(0)=&-\tfrac{1}{2}
\left(\dfrac{ma^2}{4}\right)^2\gammav(\nu+3)\cr
\times&\Bigg\{3\bigg[-\gammav(\lambda+3)\left[\tfrac{3}{2}-
\left(\dfrac{\nu+1}{2}+\psi(\nu+2)-\psi(1)\right)\right]+\dfrac{\gammav(4+\lambda)}{2}-
\phiv(\lambda,\nu)\bigg]\cr
+&2\gammav(3+\lambda)\left[\tfrac{3}{2}-\left(\dfrac{\nu+1}{2}+\psi(\nu+2)-\psi(1)\right)\right]
\left(\dfrac{\lambda+1}{2}+\psi(\lambda+2)-\psi(1)\right)\cr
-&\tfrac{1}{2}\dfrac{\gammav(\lambda+5)}{\lambda+2}
-\gammav(\lambda+4)\left[\dfrac{\lambda+1}{2}+\psi(\lambda+2)-\psi(1)\right]
+\psiv(\lambda,\nu)\Bigg\}
 R_{10}(0),
\cr}
\equn{(A3a)}$$
and the functions $\psiv,\,\phiv$ are defined as
$$\eqalign{\phiv(\lambda,\nu)=&\sum_{j=1}^{\nu+1}\dfrac{\gammav(j+\lambda+3)}{j\gammav(j+2)}\cr
\psiv(\lambda,\nu)=&\sum_{j=1}^{\nu+1}\dfrac{\gammav(j+\lambda+3)}{j\gammav(j+2)}
\left[j+\lambda+1+2\left(\psi(j+\lambda+2)-\psi(1)\right)\right]\cr
=&\sum_{j=1}^{\nu+1}\dfrac{\gammav(j+\lambda+3)}{j\gammav(j+2)}
\left[\lambda+1+2\left(\psi(j+\lambda+2)-\psi(1)\right)\right]\cr
+&\dfrac{\gammav(\lambda+\nu+5)}{(\lambda+2)\gammav(\nu+3)}
-\gammav(\lambda+2)-\gammav(\lambda+3),
\cr}
\equn{(A3b)}$$
for $\nu,\,\lambda=$~integer. The second expression for $\psiv$
 is obtained by using the identity\ref{10}
$$\sum_{k=0}^{\mu}\binom{n+k}{k}=\binom{n+\mu+1}{n+1},\equn{(A4)}$$
and has the advantage that it can be continued directly with the use of (A2).

We will require the following results:
$$\eqalign{\dfrac{\partial^2}{\partial\nu\partial\lambda}\phiv(\lambda,\nu)\Big|_{\lambda=\nu=-1}=&
\sum_{j=1}^\infty\dfrac{\psi(j+2)-j\psi'(j+2)}{j^2}\cr
=&\,\zeta_3+1-\dfrac{\pi^2\gammae}{6};\cr
\dfrac{\partial^2}{\partial\nu\partial\lambda}\psiv(\lambda,\nu)\Big|_{\lambda=\nu=-1}=&
\dfrac{\pi^2}{2}-2-\gammae\cr
-\sum_{j=1}^\infty
\Big\{2j\big[\psi''(j+1)+\psi'(j+1)\psi(j+2)\big]&
+2\left(j\psi'(j+2)-\psi(j+2)\right)\big[\psi(j+1)-\psi(1)\big]\cr
&-2\psi'(j+1)-1\Big\}/j^2.
\cr}
\equn{(A5)}$$

With the same methods we find,
$$\delta^2_{2(\lambda\nu)} R_{10}=-\tfrac{1}{4}\gammav(\nu+3)\gammav(\lambda+3)
\left(\dfrac{ma^2}{4}\right)^2\Bigg\{6\left[\tfrac{3}{2}-
\left(\dfrac{\nu+1}{2}+\psi(\nu+2)-\psi(1)\right)\right]-15+\psiv(0,\nu)\Bigg\}R_{10}(0),
\equn{(A6)} $$
and we will need $\partial\psiv(0,\nu)/\partial\nu|_{\nu=-1}$. Write
$$\psiv(0,\nu)=-3+\dfrac{(\nu+4)(\nu+3)}{2}+\nu+1+2\left[\psi(\nu+2)-\psi(1)\right]
+2\sum_{j=1}^{\nu+1}\left[\psi(j+2)-\psi(1)\right]+
4\sum_{j=1}^{\nu+1}\dfrac{\psi(j+2)-\psi(1)}{j}.
\equn{(A7a)}$$
With (cf. ref.~11 for some of the 
harmonic-type sums here\fnote{The sum $\sum_{j=1}^{\nu+1}
\left[\psi(j+2)-\psi(1)\right]$ can be evaluated 
exactly using the identity (A4) differentiated with 
respect to $n$ at $n=0$. The derivative with respect to $\nu$ of the 
sum $\sum_{j=1}^{\nu+1}\dfrac{\psi(j+2)-\psi(1)}{j}$, 
being only logarithmically divergent, can be calculated with (A2).} )
$$\eqalign{\dfrac{\partial}{\partial\nu}\sum_{j=1}^{\nu+1}
\left[\psi(j+2)-\psi(1)\right]\Big|_{\nu=-1}=&
\dfrac{\pi^2}{3}-2,\cr
\dfrac{\partial}{\partial\nu}\sum_{j=1}^{\nu+1}\dfrac{\psi(j+2)-\psi(1)}{j}\Big|_{\nu=-1}=&
\zeta_3+1\cr}
\equn{(A7b)}$$
we find
$$\dfrac{\partial\psiv(0,\nu)}{\partial\nu}=4\left(\zeta_3+1\right)+\pi^2-1/2.
\equn{(A7c)}$$

 The function
$$\bar{R}_{10}=R_{10}+\delta_{(\nu)}R_{10}+\bar{\delta}^2_{(\lambda\nu)}R_{10}$$
is not normalized to unity. The normalized wave function 
is $\hat{R}_{10}=\Vert \bar{R}_{10}\Vert^{-1/2}\bar{R}_{10}$. 
Because $R_{10}$ is orthogonal to $\delta_{\nu}R_{10}$, $\bar{\delta}^2_{\lambda\nu}R_{10}$ 
the normalization factor is
$$\Vert \bar{R}_{10}\Vert^{-1}=
\big\{1+(\delta_{(\nu)}R_{10}|\delta_{(\lambda)}R_{10})\big\}^{-1/2}\simeq1-
\tfrac{1}{2}(\delta_{(\nu)}R_{10}|\delta_{(\lambda)}R_{10}).$$
Therefore, we have to correct $\bar{\delta}^2_{(\lambda\nu)}R_{10}$ 
for this and define
$$\eqalign{\delta^2_{(\lambda\nu)}&R_{10}=
\bar{\delta}^2_{(\lambda\nu)}R_{10}+\delta^2_{\rm norm.(\lambda\nu)}R_{10},\cr
\delta^2_{\rm norm.(\lambda\nu)}&R_{10}=-\tfrac{1}{2}(\delta_{(\nu)}R_{10}|\delta_{(\lambda)}R_{10})
=-\dfrac{\gammav(\lambda+3)\gammav(\nu+3)}{2}
\left(\dfrac{ma^2}{4}\right)^2\cr
\times&\Bigg\{\tfrac{3}{4}
+\tfrac{3}{2}\left[\dfrac{\lambda+1}{2}+\psi(\lambda+2)-\psi(1)
+\dfrac{\nu+1}{2}+\psi(\nu+2)-\psi(1)\right]\cr
-&\,\left[\dfrac{\lambda+1}{2}+\psi(\lambda+2)-\psi(1)\right]
\left[\dfrac{\nu+1}{2}+\psi(\nu+2)-\psi(1)\right]\cr
-&\tfrac{1}{4}\Big[\dfrac{(\nu+1)(\nu+2)+(\lambda+1)(\lambda+2)}{2}
+5(\nu+\lambda-2)\cr
+&6\Big(\psi(\nu+2)+\psi(\lambda+2)-2\psi(1)\Big)\Big]
+\tfrac{1}{2}\sum_{j=1}^{\nu+1}\sum_{k=1}^{\lambda+1}
\dfrac{\gammav(j+k+3)}{j\gammav(j+2)k\gammav(k+2)}\Bigg\}R_{10}.
\cr}
\equn{(A8)}$$
The only difficult point is the continuation of the double sum
$$S_{\lambda\nu}=\sum_{j=1}^{\nu+1}\sum_{k=1}^{\lambda+1}
\dfrac{\gammav(j+k+3)}{j\gammav(j+2)k\gammav(k+2)}$$
which diverges if using directly (A2). 
We  avoid this by resorting again to the identity (A4).
With it, we get $S_{\lambda\nu}=S^{(1)}_{\lambda\nu}+S^{(1)}_{\nu\lambda}$,
$$\eqalign{S^{(1)}_{\lambda\nu}=&\psi(\nu+3)+\psi(\lambda+3)-2\psi(\lambda+2)-2
-2\Big[\sum_{k=2}^{\lambda+1}\dfrac{1}{k(k^2-1)}+
\sum_{k=2}^{\lambda+1}\dfrac{1}{k^2}+1\Big]+(\nu+3)\left(2+\dfrac{\nu}{4}\right)\cr
+&\dfrac{1}{\gammav(\nu+3)}\sum_{k=2}^{\lambda+1}\left[\dfrac{2\gammav(\nu+k+2)}{k-1}
+\dfrac{2\gammav(\nu+k+3)}{k}+\dfrac{2\gammav(\nu+k+4)}{k+1}\right]
\dfrac{1}{k\gammav(k+2)},\cr}$$
which can now be continued with (A2).

For our calculation, we need to evaluate
$$\left.\dfrac{\partial^2}{\partial\nu\partial\lambda}
S_{\lambda\nu}\right|_{\nu=\lambda=-1}.$$
The only nontrivial piece is the apparently divergent one 
arising for $k=2$. To get it, we note that
$$\left.\dfrac{\partial^2}{\partial\nu\partial\lambda}
\dfrac{1}{\gammav(\nu+3)}
\dfrac{2\gammav(\nu+\lambda+k+2)}{(\lambda+k)(\lambda+k-1)\gammav(\lambda+k+2}
\right|_{k=2,\nu=\lambda=-1}
=\tfrac{3}{2}\psi'(2)-\tfrac{1}{2}\psi''(2)=\pi^2/2+2\zeta_3-5. $$

It is possible also to check some of the calculations here 
in two respects. First, we may evaluate the second order correstion to the 
energy shift by $H_{1L}$.
This is given by
$$\delta^{(2)}_{1L}E_{10}=\left(\dfrac{2}{a}\right)^2 
\dfrac{C_F^2\beta_0^2\alpha_s^4}{4\pi^2}
\dfrac{\partial^2}{\partial\nu\partial\lambda}(R_{10},\rho^\lambda P_{10}
\dfrac{1}{E^{(0)}-H^{(0)}} P_{10}\rho^\nu R_{10})
$$
Using our formulas here we find
$$-m\frac{C_F^2\beta_0^2\alpha_s^4}{4\pi^2}\dfrac{3+3\gamma_E^2-\pi^2+6\zeta_3}{12},$$
in agreement with the result of ref.~5, obtained with
 a completely diferent method (Green's functions).
Secondly, we can particilarize the calculation here 
for $\nu=\lambda=-1$. We get
$$\delta^2_{(-1,-1)}R_{10}=\left(\tfrac{3}{4}-\tfrac{3}{4}\rho+\tfrac{1}{8}\rho^2\right)
\left(\dfrac{ma^2}{4}\right)^2R_{10}-\tfrac{3}{8}R_{10}$$
and the last term is the normalization correction. This is 
in  agreement with the direct result
$$\delta^2_{(-1,-1)}R_{10}=\left(\tfrac{3}{8}-\tfrac{3}{4}\rho+\tfrac{1}{8}\rho^2\right)
\left(\dfrac{ma^2}{4}\right)^2.$$

With these evaluations we can get the result of the second order 
perturbation by $\rho^{-1}\log\rho$. 
Write
$$\delta^{(2)}_{(\rho^{-1}\log\rho)}R_{10}=
\dfrac{\partial^2}{\partial\lambda\partial\nu}\delta^2_{(\lambda\nu)}R_{10}\equiv
 \left(\dfrac{ma^2}{4}\right)^2R_{10}\left[d_1-d_2+d_{\rm norm.}\right].$$

One defines
$$\eqalign{N_0\equiv&\dfrac{\partial^2}{\partial\lambda\partial\nu}S_{\lambda\nu}\big|_{\lambda=\nu=-1}
=\dfrac{\pi^2}{2}+2\zeta_3-5\cr
+&2\sum_{k=1}^\infty\Bigg\{2\dfrac{\psi(k+2)-\psi(2)}{k(k+1)(k+2)}
\left[\dfrac{1}{k}+\dfrac{1}{k+1}+\dfrac{1}{k+2}\right]
+4\dfrac{\psi(k+2)-\psi(2)}{k^3}\cr
+&\dfrac{\psi(k+3)-\psi(2)}{k(k+1)}
\left[\dfrac{k+2}{k}+\dfrac{1}{k+1}\right]
-2\dfrac{\psi'(k+2)}{k(k+1)(k+2)}-2\dfrac{\psi'(k+2)}{k^2}-
\dfrac{(k+2)\psi'(k+3)}{k(k+1)}\Bigg\}\cr
\simeq&10.29;\cr
N_1\equiv&
\dfrac{\partial^2}{\partial\lambda\partial\nu}\psiv(\lambda,\nu)\Big|_{\lambda=\nu=-1}
\simeq6.82\cr}$$
(the exact value of the last given in (A5)).
Moreover,
$$
\dfrac{\partial^2}{\partial\lambda\partial\nu}\phiv(\lambda,\nu)\Big|_{\lambda=\nu=-1}=
\sum_{j=1}^\infty\dfrac{\psi(j+2)-j\psi'(j+2)}{j^2}=\zeta_3+1-\dfrac{\pi^2\gammae}{6}.
\equn{(A9a)}$$
Some of the sums in the numbers $N_i$  are simplified  
in terms of the following harmonic sums\ref{11}:
$$\eqalign{\sum_{j=1}^\infty\dfrac{1}{j^2}\sum_{k=1}^j\dfrac{1}{k}=2\zeta_3&;\qquad
\sum_{j=1}^\infty\dfrac{1}{j^3}\sum_{k=1}^j\dfrac{1}{k}=\pi^4/18;\cr
\sum_{j=1}^\infty\dfrac{1}{j}\sum_{k=1}^\infty\dfrac{1}{(k+j)^2}=\zeta_3&;\qquad
\sum_{j=1}^\infty\dfrac{1}{j^2}\sum_{k=1}^\infty\dfrac{1}{(k+j)^2}=\pi^4/120.\cr}$$ 
Then,
$$\eqalign{d_1=&-\tfrac{1}{2}\Bigg\{(1-\gammae)
\left[\dfrac{2}{3}\pi^2+6\gammae-\tfrac{11}{4}\right]
-\tfrac{1}{2}\left(1+\dfrac{\pi^2}{3}\right)^2
-3\left[\zeta_3+1-\dfrac{\pi^2\gammae}{6}\right]+N_1\Bigg\};\cr
d_2=&-\tfrac{1}{4}(1-\gammae)\left\{-\tfrac{19}{2}+6\gammae+4\zeta_3\right\};\cr
d_{\rm norm.}=&-\tfrac{3}{8}(1-\gammae)^2
+\tfrac{5}{8}(1-\gammae)+
\tfrac{1}{2}\left(\tfrac{1}{2}+\dfrac{\pi^2}{6}\right)^2-\tfrac{1}{4}N_0.\cr}.
\equn{(A9b)}$$
Finally, the coefficient $c_{L,1}^{(2)}$ is
$$c_{L,1}^{(2)}=d_1-d_2+d_{\rm norm.}\simeq1.75.
\equn{(A9c)}$$

\brochuresection{ Acknowledgments}
 I am grateful to Drs. A.~Pineda, J.~A. M. Vermaseren and T.~van~Ritbergen 
for encouragement and 
discussions. Thanks are due to CICYT, Spain, for financial support.

\vfill\eject
\brochuresection{References}
\item{1.}{\ajnyp{R. Coquereaux}{Phys. Rev.}{D23}{1981}{1365}; 
\ajnyp{R. Tarrach}{Nucl. Phys.}{B183}{1981}{384}; 
\ajnyp{N.~Gray et al.}{Z. Phys.}{C48}{1990}{673};
\ajnyp{K.~Melnikov and T.~van~Ritbergen}{}{}{hep-ph/9912391}.}
\item{2. }{\ajnyp{N. Brambilla et al.,}{Phys. Lett.}{470B}{1999}{215} for 
the $\alpha_s^5\log \alpha_s$ correction; 
{\sc F.~J.~Yndur\'ain}, FTUAM-00-07, 2000
(hep-ph/0002237) for the $m_c^2/m_b^2$ one.}
\item{3. }{\ajnyp{R. Barbieri et al.,}{Phys. Lett.}{57B}{1975}{455}; ibid. 
{\sl Nucl. Phys.} {\bf B 154} (1979) 535, to 
one loop, and \ajnyp{M. Beneke, A. Signer and V. A. Smirnov}{Phys. Rev. Lett.}{80}{1998}{2535} 
for the two loop evaluation of the ``hard" 
part of the correction. 
The two loop correction to the wave function, 
in the context of heavy quark production 
by $e^+e^-$, has been evaluated by
 \ajnyp{A.~A. Penin and A.~A.~Pivovarov}{Nucl. Phys.}{B549}{1999}{217};
\ajnyp{K. Melnikov and A. Yelkhovsky}{Phys. Rev.}{D59}{1999}{114009}.} 
\item{4.}{\ajnyp{S. Titard and F. J. Yndur\'ain}{Phys. Rev.}{D49}{1994}{6007}.}
\item{5.}{\ajnyp{A. Pineda and F. J. Yndur\'ain}{Phys. Rev.}{D58}{1998}{3003}, and 
 {\bf D61} (2000) 
077505.}
\item{6.}{\ajnyp{M. Beneke and A. Signer}{Phys. Lett.}{B471}{1999}{233}} 
\item{7.}{See \ajnyp{J.~Santiago and F.~J.~Yndur\'ain} 
{Nucl. Phys.}{B563}{1999}{45}, and work quoted there.}
\item{8.}{\ajnyp{W. Lucha and F. Sch\"oberl}{UWThPh}{}{1999}{233} 
(hep-ph/0001191).} 
\item{9.}{\ajnyp{M. B. Voloshin}{Nucl. Phys.}{B154}{1979}{365} and 
{\sl Sov. J. Nucl. Phys.} 
{\bf 36} (1982) 143; \ajnyp{H. Leutwyler}{Phys. Lett.}{B98}{1981}{447}.}
\item{10. }{{\sc I. S. Gradshteyn and I. M. Ryzhik, }{\sl Tables of 
Integrals, Series and Functions} (5 edition), {Academic Press}, New York 1994, 
\equs.~KR64(70.1), p. 4.}
\item{11. }{\ajnyp{J. Vermaseren}{Int. J. Mod. Phys.}{A14}{1999}{41}.}

\bye